\documentclass[12pt]{JHEP3}
\usepackage{graphicx}
\usepackage{amsmath}

\normalbaselines

\title{Reduction Schemes in Cutoff Regularization and Higgs Decay into Two Photons}

\author{Hua-Sheng Shao\\
Department of Physics and State Key Laboratory of Nuclear Physics
and Technology, Peking University,
 Beijing 100871, China\\
 Email:\email{erdissshaw@gmail.com}}

\author{Yu-Jie Zhang\\
Key Laboratory of Micro-nano
 Measurement-Manipulation and Physics (Ministry of Education) and School of Physics, Beihang University,
 Beijing 100191, China\\
 Email:\email{nophy0@gmail.com}}

\author{Kuang-Ta Chao\\
 Department of Physics and State Key Laboratory of
Nuclear Physics and Technology, Peking University,
 Beijing 100871, China\\
Center for High Energy Physics, Peking University, Beijing 100871,
China\\
Email:\email{ktchao@pku.edu.cn}}

\abstract{We present a new systematic method to evaluate one-loop
tensor integrals in conventional ultraviolet cutoff regularization.
By deriving a new recursive relation that describes the momentum
translation variance of ultraviolet integrals, we implement this
relation in the Passarino-Veltman reduction method. With this
method, we recalculated the Higgs boson decay into two photons
process at one-loop level in the Standard Model. We reanalyze this
process carefully and clarify some issues arisen recently in cutoff
regularization.}


\begin{document}

\section{Introduction}
Recently, the ATLAS\cite{ATLAS:2011} and CMS\cite{CMS:2011}
collaborations have renewed efforts to search for the Higgs boson at
the CERN LHC with data integrated  up to $\mathcal{O}(fb^{-1})$. The
excluded mass region for the Standard Model(SM) Higgs boson has been
extended to most of the region between 145 and 466 GeV. In the low
mass region of the Higgs boson, the two-photon mode of Higgs decay
plays a crucial role in experimental studies.

R. Gastmans {\it et al.} recently recalculated $H\rightarrow
\gamma\gamma$  via W-boson loop
\cite{Gastmans:2011ks,Gastmans:2011wh}, which yielded a result in
contradiction with the old ones in the
literature\cite{Ellis:1975ap,Ioffe:1976sd,Shifman:1979eb,Rizzo:1979mf}.
Their computation was carried out in four-momentum cutoff
regularization rather than dimensional regularization (DREG). To
reduce the number of Feynman diagrams, Gastmans {\it et al.} chose
unitary gauge. In their treatment, the new result, which satisfies
the decoupling theorem\cite{Appelquist:1974tg}, was favored by the
authors. Later, several
authors\cite{Shifman:2011ri,Huang:2011yf,Marciano:2011,Jegerlehner:2011jm}
have pointed out that the old results are still correct and
decoupling theorem is violated by $H\phi^+\phi^-$ in this case.
However, we are still unsatisfactory with the explanations about the
problems with the calculations of R.Gastmans {\it et al.}, since
there have never been doubts about the correctness of their algebra.
In order to clarify this problem, we develop a new method to do
one-loop calculation in cutoff regularization.

Although DREG has proven its superiority and achieved the most
widely usage in phenomenological applications, cutoff
regularization, the oldest regularization, still has some advantages
compared with DREG theoretically. For instance, in DREG, one is
unable to obtain the correct divergent terms higher than logarithmic
divergences, which means that quadratic divergent terms of SM Higgs
self-energy diagrams disappeared in DREG. Pauli-Villars
regularization is flawed because it violates chiral symmetry, while
the symmetry is preserved in cutoff regularization. Moreover, from
Wilsonian effective field theory viewpoint, cutoff regularization
scheme is also a more intuitive and straightforward scheme.
Therefore, the introduction of an explicit cutoff is sometimes
advantageous.

However, there are still many drawbacks in this four-momentum
regularization that should be mentioned.Considering the truncation
in momentum modes, this regularization is flawed because it violates
gauge invariance and translation invariance regarding the loop
momentum. The latter condition signifies that the results may
ambiguously depend on the manner how the propagators are written.
Hence, in the present paper a new recursive relation for loop
momentum translation is derived first. Then the Passarino-Veltman
reduction method{\footnote{Note that, integration by parts (IBP)
reduction methods are not valid in this case due to nonvanishing
surface terms.}}\cite{Passarino:1978jh,Denner:2005nn} is modified to
reduce the tensor integrals in this regularization. One can follow
Dyson's prescription\cite{Dyson:1949bp,Dyson:1949ha} to obtain a
gauge invariant result, just as shown in our calculations of
$H\rightarrow \gamma\gamma$.

As an example, we reconsider the process $H\to \gamma\gamma$ in this
four-dimensional momentum cutoff regularization with our proposed
approach. Given that the process of $H\rightarrow \gamma\gamma$ is
free from infrared and mass singularities, only the ultraviolet
cutoff is considered here. Readers who need to handle infrared or
mass singularities should turn to the mass regularization scheme
demonstrated in the literature e.g.\cite{Denner:2010tr}.

The present paper is organized as follows. In Section \ref{sec:2}, a
new recursive relation for the loop momentum translation in cutoff
regularization is demonstrated. Then, it is implemented in the
Passarino-Veltman reduction schemes in Section \ref{sec:3}. With
this approach, calculations and analysis of
$H\rightarrow\gamma\gamma$ are performed in Section \ref{sec:4}. Our
conclusion is present in Section \ref{sec:5}. In
Appendex\ref{app:a}, the expressions for $J^N_{\mu_1\ldots\mu_s}$
used in Section \ref{sec:2} is derived. Finally, some scalar
integrals can be found in Appendix \ref{app:b}.
\section{A New Recursive Relation \label{sec:2}}
In this section, we will show how to calculate
\begin{center}
\begin{eqnarray}
I^{\Delta}_{\mu_1\ldots\mu_s}(b,a^2)\equiv\int{{\rm
d}^4k\frac{k_{\mu_1}\ldots k_{\mu_s}}{(-(k-b)^2+a^2)^n}}-\int{{\rm
d}^4k\frac{(k+b)_{\mu_1}\ldots (k+b)_{\mu_s}}{(-k^2+a^2)^n}}.
\label{eq:In}
\end{eqnarray}
\end{center}
This integration have a superficial divergence degree $\Delta\equiv
s+4-2n$.

A negative $\Delta$ evidently simplifies the calculation, because
the limits of the integrals in Eq.(\ref{eq:In}) can be set to
infinity and the translation shift $k\rightarrow k+b$ does not
change these limits. Therefore, $I^{\Delta}_{\mu_1\ldots\mu_s}$
completely vanishes when $\Delta<0$. However, results may vary when
the integrals in Eq.(\ref{eq:In}) are ultraviolet divergent because
there is an artificial four-momentum cutoff scale $\mathbf{\Lambda}$
in these integrals. These conditions are then considered following.

$I^{\Delta}_{\mu_1\ldots\mu_s}$ can be rewritten as
\begin{center}
\begin{eqnarray}
I^{\Delta}_{\mu_1\ldots\mu_s}(b,a^2)&=&\left(\int{{\rm
d}^4k\frac{k_{\mu_1}\ldots k_{\mu_s}}{(-(k-b)^2+a^2)^n}}-\int{{\rm
d}^4k\frac{k_{\mu_1}\ldots
k_{\mu_s}}{(-k^2+a^2)^n}}\right)\nonumber\\
&&-\int{{\rm d}^4k\frac{f^{rem}(b)}{(-k^2+a^2)^n}},
\end{eqnarray}
\end{center}
with the remainder $f^{rem}(b)\equiv(k+b)_{\mu_1}\ldots
(k+b)_{\mu_s}-k_{\mu_1}\ldots k_{\mu_s}$. Using the identity
\begin{center}
\begin{eqnarray}
\frac{1}{A^n}-\frac{1}{B^n}=\int^{1}_{0}{{\rm d}x\frac{n(B-A)}{(x
A+(1-x)B)^{n+1}}},
\end{eqnarray}
\end{center}
one arrived at
\begin{center}
\begin{eqnarray}
I^{\Delta}_{\mu_1\ldots\mu_s}(b,a^2)&=&\int{{\rm
d}^4k\int^{1}_{0}{{\rm d}x \frac{n(-2b\cdot k+b^2)k_{\mu_1}\ldots
k_{\mu_s}}{(-(k-c)^2+d^2)^{n+1}}}}\nonumber\\
&&-\int{{\rm d}^4k\frac{f^{rem}(b)}{(-k^2+a^2)^n}},
\end{eqnarray}
\end{center}
where $c\equiv x~b,d^2\equiv a^2-b^2x(1-x)$.  The integral momentum
$k$ to $k+c$ in the first integral of the previous equation is
shifted, so that
\begin{center}
\begin{eqnarray}
I^{\Delta}_{\mu_1\ldots\mu_s}(b,a^2)&=&-2n~b^{\mu_0}\int^{1}_{0}{{\rm
d}x I^{\Delta-1}_{\mu_0\mu_1\ldots
\mu_s}(c,d^2)}+n~b^2\int^{1}_{0}{{\rm d}x I^{\Delta-2}_{\mu_1\ldots
\mu_s}(c,d^2)}\nonumber\\
&&+\int{{\rm d}^4k\int_{0}^{1}{{\rm d}x\frac{n(-2b\cdot
k+b^2-2b\cdot
c)k_{\mu_1}\ldots k_{\mu_s}}{(-k^2+d^2)^{n+1}}}}\nonumber\\
&&+\int{{\rm d}^4k\int^{1}_{0}{{\rm d}x\frac{n(-2b\cdot
k+b^2-2b\cdot
c)f^{rem}(c)}{(-k^2+d^2)^{n+1}}}}\nonumber\\
&&-\int{{\rm d}^4k\frac{f^{rem}(b)}{(-k^2+a^2)^n}}\nonumber
\end{eqnarray}
\end{center}
\begin{center}
\begin{eqnarray}
&=&-2n~b^{\mu_0}\int^{1}_{0}{{\rm d}x I^{\Delta-1}_{\mu_0\mu_1\ldots
\mu_s}(c,d^2)}+n~b^2\int^{1}_{0}{{\rm d}x I^{\Delta-2}_{\mu_1\ldots
\mu_s}(c,d^2)}\nonumber\\
&&-2n~b^{\mu_0}\int{{\rm d}^4k\int^{1}_{0}{{\rm
d}x\frac{k_{\mu_0}k_{\mu_1}\ldots
k_{\mu_s}}{(-k^2+d^2)^{n+1}}}}\nonumber\\
&&-2n~\int{{\rm d}^4k\int^{1}_{0}{{\rm d}x\frac{b\cdot k
f^{rem}(c)}{(-k^2+d^2)^{n+1}}}}\nonumber\\
&&-\int{{\rm d}^4k\int^{1}_{0}{{\rm d}x\frac{\frac{\partial
f^{rem}(c)}{\partial x}}{(-k^2+d^2)^n}}}.\label{eq:In2}
\end{eqnarray}
\end{center}
In the sixth line of Eq.(\ref{eq:In2}), a spurious part that is
proportional to $(1-2x)$ in the numerator of the integrand is
removed. What's more, integration by parts is performed at the end
of Eq.(\ref{eq:In2}).

Aside from the terms expressed in $I^{\Delta-1}$ and $I^{\Delta-2}$,
Eq.(\ref{eq:In2}) can be simplified further using the formulae
$J^N_{\mu_1\ldots \mu_s}(a^2)\equiv\int{{\rm
d}^4k\frac{k_{\mu_1}k_{\mu_2}...k_{\mu_s}}{(-k^2+a^2)^N}}$ given in
appendix \ref{app:a}. After expanding the terms proportional to
$x^j$ in the integrands and implementing the expressions for $J^N$,
a lot of terms are canceled. Thus, the final result  is
\begin{center}
\begin{eqnarray}
I^{\Delta}_{\mu_1\ldots
\mu_s}(b,a^2)&=&-2n~b^{\mu_0}\int^{1}_{0}{{\rm d}x
I^{\Delta-1}_{\mu_0\mu_1\ldots \mu_s}(c,d^2)}+n~b^2\int^{1}_{0}{{\rm
d}x I^{\Delta-2}_{\mu_1\ldots
\mu_s}(c,d^2)}\nonumber\\
&&-\sum^{2\lfloor\frac{s+1}{2}\rfloor-2n+2}_{t=\max(0,4-2n),~\text{even}}{\{g^{s+t-\Delta}b^{\Delta-t}\}_{\mu_1\ldots
\mu_s}\frac{i\pi^2(-2)^{-\frac{s+t-\Delta}{2}}}{\Gamma(\frac{s+t-\Delta}{2}+3)}h(t,n,\Delta)},\nonumber\\~\label{eq:In3}
\end{eqnarray}
\end{center}
where the notation $\{g^{s+t-\Delta}b^{\Delta-t}\}_{\mu_1\ldots
\mu_s}$  defined in appendix \ref{app:a}, $n=\frac{s+4-\Delta}{2}$,
$ \lfloor y\rfloor$ is a Gaussian function ( the greatest integer
that is not larger than $y$), and the function is
\begin{center}
\begin{eqnarray}
h(t,n,\Delta)\equiv\sum^{\frac{t}{2}}_{l=0}{C_{n-1+l}^l\sum_{n_2+n_1+n_0=l}^{n_2,n_1,n_0\geq0}{
\frac{(-1)^{n_2+n_0}(\Delta-t)}{2n_2+n_1+\Delta-t}\frac{l!}{n_2!n_1!n_0!}(b^2)^{n_2+n_1}(a^2)^{n_0}\Lambda^{t-2l}}}.\nonumber\\
~
\end{eqnarray}
\end{center}
We should make some remarks about above equation before going
forward.In the recursive relation Eq. (\ref{eq:In3}), there are two
integrals left. However, since all of the $I^{\Delta}_{\mu_1\ldots
\mu_s}(b,a^2)$ are only polynomials of $b$ and $a^2$,i.e.,they are
only polynomials of integral variable $x$, the explicit expressions
for this recursive relation can be easily obtained with the help of
computers. Especially, the explicit expressions for
$I^{\Delta}_{\mu_1\ldots\mu_s} (\Delta=0,1,2,3)$ are
\begin{center}
\begin{eqnarray}
I^{0}_{\mu_1\ldots\mu_s}(b,a^2)&=&0,\nonumber\\
I^{1}_{\mu_1\ldots\mu_s}(b,a^2)&=&-\{g^{s-1}b^1\}_{\mu_1\ldots\mu_s}
\frac{i\pi^2(-2)^{-\frac{s-1}{2}}}{\Gamma(\frac{s+5}{2})},\nonumber\\
I^{2}_{\mu_1\ldots\mu_s}(b,a^2)&=&
\left(n~b^2\{g^s\}_{\mu_1\ldots\mu_s}+(4n+2)\theta(s-2)\{g^{s-2}b^2\}_{\mu_1\ldots\mu_s}\right)\nonumber\\
&&\frac{i\pi^2(-2)^{1-n}}{\Gamma(n+2)},~~~~~~~~~~~\text{with}~~~n=\frac{s+2}{2},\nonumber
\end{eqnarray}
\end{center}
\begin{center}
\begin{eqnarray}
I^{3}_{\mu_1\ldots\mu_s}(b,a^2)&=&-\theta(s-3)\{g^{s-3}b^3\}_{\mu_1\ldots\mu_s}
\frac{i\pi^2(-2)^{2-n}(3n^2+6n+2)}{\Gamma(n+3)}\nonumber\\
&&-\{g^{s-1}b^1\}_{\mu_1\ldots\mu_s}\frac{i\pi^2(-2)^{1-n}}{\Gamma(n+2)}\nonumber\\
&&\left(\Lambda^2-n~a^2-\frac{n(n+1)}{n+2}b^2\right),~~~\text{with}~~~n=\frac{s+1}{2}.
\end{eqnarray}
\end{center}

\section{Modified Passarino-Veltman Reduction Schemes\label{sec:3}}
It is known that the one-loop tensor integrals can be reduced to a
linear combination of up to four-point scalar
integrals\cite{Passarino:1978jh}. In this section, a generic
one-loop integral
\begin{center}
\begin{eqnarray}
T^N_{\mu_1\ldots\mu_s}&\equiv&\int{{\rm d}^4k\frac{k_{\mu_1}\ldots
k_{\mu_s}}{D_0D_1\ldots D_{N-1}}},
\end{eqnarray}
\end{center}
with propagators $D_i\equiv(k+p_i)^2-m_i^2+i\varepsilon$ and $p_0=0$
is considered. As mentioned in the previous sections, the integrals
$T^N_{\mu_1\ldots\mu_s}$ may not be translation invariant because of
the finite integral limits when they are ultraviolet divergent.
Therefore, the expressions for
\begin{center}
\begin{eqnarray}
\Delta L^N_{\mu_1\ldots\mu_s}&\equiv&\int{{\rm
d}^4k\frac{k_{\mu_1}\ldots k_{\mu_s}}{D_1\ldots D_{N}}}-\int{{\rm
d}^4k\frac{(k-p_1)_{\mu_1}\ldots (k-p_1)_{\mu_s}}{\tilde{D}_1\ldots
\tilde{D}_N}},
\end{eqnarray}
\end{center}
where propagators $D_i\equiv(k+p_i)^2-m_i^2+i\varepsilon$ but
$p_1\neq0$ and $\tilde{D}_i\equiv(k+p_i-p_1)^2-m_i^2+i\varepsilon$
should be calculated. After the conventional Feynman
parameterization, $\Delta L^N_{\mu_1\ldots \mu_s}$ can be
reexpressed as
\begin{center}
\begin{eqnarray}
\Delta
L^N_{\mu_1\ldots\mu_s}&\equiv&(-)^N\Gamma(N)\left[\int_{\text{simplex}}{\prod^N_{i=1}{{\rm
d}u_i}~I^{s+4-2N}_{\mu_1\ldots\mu_s}(\tilde{b}-p_1,\tilde{a}^2)}\right.\nonumber\\
&&\left.-\int_{\text{simplex}}{\prod^N_{i=1}{{\rm
d}u_i}~\sum^{s}_{i=0}{(-)^{s-i}\{p_1^{s-i}I^{i+4-2N,~i}(\tilde{b},\tilde{a}^2)\}_{\mu_1\ldots
\mu_s}}}\right],\label{eq:Ln}
\end{eqnarray}
\end{center}
where
$\tilde{b}\equiv-\sum^{N}_{i=1}{u_ip_i}+p_1,\tilde{a}^2\equiv-\sum^N_{i=1}{u_i(p_i^2-m_i^2)
+\sum^{N}_{i,j=1}{u_iu_jp_i\cdot p_j}}$, and the notation
$\{p_1^{s-i}I^{i+4-2N,i}(\tilde{b},\tilde{a}^2)\}_{\mu_1\ldots
\mu_s}$ is defined in appendix \ref{app:a} with the divergence
degree of $I^{i+4-2N,~i}$ defined in Section \ref{sec:2} is
$i+4-2N$. The polynomial dependence of $b,a^2$ in
$I^{\Delta}(b,a^2)_{\mu_1\ldots \mu_s}$ makes the simplex
integration in Eq.(\ref{eq:Ln}) straightforward using
\begin{center}
\begin{eqnarray}
\int_{\text{simplex}}{\prod^N_{i=1}{{\rm
d}u_i}\prod^N_{i=1}{u_i^{r_i-1}}}=\frac{\prod^N_{i=1}{\Gamma(r_i)}}{\Gamma(\sum^N_{i=1}{r_i})}.
\end{eqnarray}
\end{center}

Next, several notations similar to that given in
ref.\cite{Denner:2005nn} are reintroduced here
\begin{center}
\begin{eqnarray}
\Delta
L^N_{\mu_1\ldots\mu_s}&\equiv&\sum^{n_0,n_1,\ldots,n_N\geq0}_{2n_0+n_1+\ldots+n_N=s}{\{g^{2n_0}p_1^{n_1}\ldots
p_N^{n_N}\}_{\mu_1\ldots\mu_s}\Delta L^N_{{\scriptsize
\underbrace{0\ldots0}_{2n_0}\ldots\underbrace{N\ldots
N}_{n_N}}}}, \nonumber\\
T^N_{\mu_1\ldots\mu_s}&\equiv&\sum^{n_0,n_1,\ldots,n_{N-1}\geq0}_{2n_0+n_1+\ldots+n_{N-1}=s}{\{g^{2n_0}p_1^{n_1}\ldots
p_{N-1}^{n_{N-1}}\}_{\mu_1\ldots\mu_s} T^N_{{\scriptsize
\underbrace{0\ldots0}_{2n_0}\ldots\underbrace{(N-1)\ldots
(N-1)}_{n_{N-1}}}}},\nonumber\\
T^N_{\mu_1\ldots\mu_s}(0)&\equiv&\int{{\rm
d}^4k\frac{k_{\mu_1}\ldots k_{\mu_s}}{D_1\ldots D_{N}}},\nonumber\\
T^N_{\mu_1\ldots\mu_s}(k)&\equiv&\int{{\rm
d}^4k\frac{k_{\mu_1}\ldots k_{\mu_s}}{D_0\ldots \hat{D}_{k}\ldots D_{N}}},\nonumber\\
\tilde{T}^N_{\mu_1\ldots\mu_s}(0)&\equiv&\int{{\rm
d}^4k\frac{k_{\mu_1}\ldots
k_{\mu_s}}{\tilde{D}_1\ldots\tilde{D}_N}},\nonumber\\
T^N_{\mu_1\ldots\mu_s}(0)&\equiv&\sum^{n_0,n_1,\ldots,n_N\geq0}_{2n_0+n_1+\ldots+n_N=s}{\{g^{2n_0}p_1^{n_1}\ldots
p_N^{n_N}\}_{\mu_1\ldots\mu_s} T^N_{{\scriptsize
\underbrace{0\ldots0}_{2n_0}\ldots\underbrace{N\ldots
N}_{n_N}}}(0)},\nonumber\\
\tilde{T}^N_{\mu_1\ldots\mu_s}(0)&\equiv&\sum^{n_0,n_1,\ldots,n_{N-1}\geq0}_{2n_0+n_1+\ldots+n_{N-1}=s}
{\{g^{2n_0}(p_2-p_1)^{n_1}\ldots
(p_N-p_1)^{n_{N-1}}\}_{\mu_1\ldots\mu_s}}\nonumber\\
&&{\tilde{T}^N_{{\scriptsize
\underbrace{0\ldots0}_{2n_0}\ldots\underbrace{(N-1)\ldots
(N-1)}_{n_{N-1}}}}(0)},
\end{eqnarray}
\end{center}
with $D_0\ldots \hat{D}_k \ldots D_N\equiv D_0\ldots
D_{k-1}D_{k+1}\ldots D_N$ and employing the caret "\verb"^"" to
indicate the indices omitted.

Thus in this cutoff regularization, Eq.(2.9) in
ref.\cite{Denner:2005nn} should be replaced by
\begin{center}
\begin{eqnarray}
T^N_{{\scriptsize
\underbrace{0\ldots0}_{2n}\underbrace{1\ldots1}_k}i_{2n+k+1}\ldots
i_s}(0)&=&(-)^k\sum^{k}_{l=0}{C^l_k\sum^{N-1}_{i_1,\ldots,i_l=1}{\tilde{T}^N_{{\scriptsize
\underbrace{0,\ldots,0}_{2n},i_1,\ldots,
i_l,~i_{2n+k+1}-1,\ldots,~i_s-1}}(0)}}\nonumber\\
&&+\Delta L^N_{{\scriptsize
\underbrace{0\ldots0}_{2n}\underbrace{1\ldots1}_k}i_{2n+k+1}\ldots
i_s},~~i_{2n+k+1},\ldots,i_s>1.
\end{eqnarray}
\end{center}

When the determinant of Gram matrix
\begin{center}
\begin{eqnarray}
Z^{(N)}&=&\left(\begin{array}{c c c} 2p_1\cdot p_1&\ldots& 2p_1\cdot
p_N\\ \vdots & \ddots & \vdots\\ 2p_N\cdot p_1 & \ldots & 2p_N\cdot
p_N
\end{array}\right)
\end{eqnarray}
\end{center}
for $(N+1)$-point functions is non-vanishing, the reduction can be
continued as
\begin{center}
\begin{eqnarray}
T^N_{00i_3\ldots i_s}&=&\frac{1}{2(3+s-N)}\left[T^{N-1}_{i_3\ldots
i_s}(0)+2m_0^2T^N_{i_3\ldots
i_s}+\sum^{N-1}_{j=1}{f_jT^N_{ji_3\ldots
i_s}}\right.\nonumber\\
&&\left.+\Delta L^{N-1}_{i_3\ldots i_s}+\sum^{N-1}_{j=1}{\Delta
L^{N-1}_{ji_3\ldots i_s}}\right],\nonumber\\
T^N_{i_1\ldots
i_s}&=&\sum^{N-1}_{j=1}{(Z^{(N-1)})^{-1}_{i_1j}\left(S^s_{ji_2\ldots
i_s}-2\sum^s_{r=2}{\delta_{ji_r}T^N_{00i_2\ldots \hat{i}_r \ldots
i_s}}\right)},~~i_1\neq0,
\end{eqnarray}
\end{center}
where some notations are defined in ref.\cite{Denner:2005nn}
\begin{center}
\begin{eqnarray}
f_k&\equiv&p_k^2-m_k^2+m_0^2,\nonumber\\
\bar{\delta}_{ij}&\equiv&1-\delta_{ij},\nonumber\\
(i_r)_k&\equiv&\left\{\begin{array}{c@{\:,\:}l}i_r&k>i_r\\i_r-1&k<i_r\end{array}\right.,\nonumber\\
S^s_{ki_2\ldots i_s}&\equiv&T^{N-1}_{(i_2)_k\ldots
(i_s)_k}(k)\bar{\delta}_{ki_2}\ldots
\bar{\delta}_{ki_s}-T^{N-1}_{i_2\ldots i_s}(0)-f_kT^N_{i_2\ldots
i_s}.
\end{eqnarray}
\end{center}
Otherwise, when the Gram determinant is zero, there is at least
one non-vanishing element $\tilde{Z}^{(N)}_{kl}$ in the adjoint
matrix of $Z^{(N)}$
\begin{center}
\begin{eqnarray}
\tilde{Z}^{(N)}_{kl}&\equiv&(-)^{k+l}\left|\begin{array}{*{6}{c}}2p_1p_1&\ldots&2p_1p_{l-1}&2p_1p_{l+1}
&\ldots&2p_1p_N\\\vdots&\ddots&\vdots&\vdots&\ddots&\vdots\\2p_{k-1}p_1&\ldots&2p_{k-1}p_{l-1}&
2p_{k-1}p_{l+1}&\ldots&2p_{k-1}p_N\\2p_{k+1}p_1&\ldots&2p_{k+1}p_{l-1}&2p_{k+1}p_{l+1}&\ldots&
2p_{k+1}p_N\\\vdots&\ddots&\vdots&\vdots&\ddots&\vdots\\2p_Np_1&\ldots&2p_Np_{l-1}&2p_Np_{l+1}
&\ldots&2p_Np_N\end{array}\right|
\end{eqnarray}
\end{center}
and one non-zero element $\tilde{X}^{(N)}_{0j}$ in the adjoint
matrix of the following matrix
\begin{center}
\begin{eqnarray}
X^{(N)}&\equiv&\left(\begin{array}{*{4}{c}}2m_0^2&f_1&\ldots&f_N\\f_1&2p_1p_1&\ldots&2p_1p_N\\\vdots&
\vdots&\ddots&\vdots\\f_N&2p_Np_1&\ldots&2p_Np_N\end{array}\right).
\end{eqnarray}
\end{center}
One-loop reduction can be applied using the following equations
\begin{center}
\begin{eqnarray}
T^N_{i_1\ldots
i_s}&=&-\frac{1}{\tilde{X}^{(N-1)}_{0j}}\sum^{N-1}_{n=1}{\tilde{Z}^{(N-1)}_{jn}\left(\hat{S}^{s+1}_{ni_1\ldots
i_s}- 2\sum^s_{r=1}{\delta_{ni_r}T^N_{00i_1\ldots \hat{i}_r\ldots
i_s}}\right)},\nonumber\\
T^N_{00i_1\ldots
i_s}&=&\frac{1}{2(6+s-N+\sum^s_{r=1}{\bar{\delta}_{i_r0}})\tilde{Z}^{(N-1)}_{kl}}\left[\tilde{Z}^{(N-1)}_{kl}
S^{s+2}_{00i_1\ldots i_s}\right.\nonumber\\
&&+\sum^{N-1}_{n=1}{\left(\tilde{Z}^{(N-1)}_{nl}\hat{S}^{s+2}_{nki_1\ldots
i_s}-\tilde{Z}^{(N-1)}_{kl}\hat{S}^{s+2}_{nni_1\ldots
i_s}\right)}\nonumber\\&&-\sum^{N-1}_{n,m=1}{\tilde{\tilde{Z}}^{(N-1)}_{(kn)(lm)}
\left(f_n\hat{S}^{s+1}_{mi_1\ldots
i_s}+2\sum^s_{r=1}{\delta_{ni_r}\hat{S}^{s+2}_{m00i_1\ldots
\hat{i}_r\ldots i_s}}\right.}\nonumber\\
&&-f_nf_mT^{N}_{i_1\ldots
i_s}-2\sum^s_{r=1}{\left(f_n\delta_{mi_r}+f_m\delta_{ni_r}\right)T^{N}_{00i_1\ldots
\hat{i}_r\ldots i_s}}\nonumber\\&&\left.{\left.-4\sum^s_{r,t=1,r\neq
t}{\delta_{ni_r}\delta_{mi_t}T^{N}_{0000i_1\ldots\hat{i}_r\ldots\hat{i}_t\ldots
i_s}}\right)}\right].\label{eq:Tn1}
\end{eqnarray}
\end{center}
Some notations in Eq.(\ref{eq:Tn1}) should be recalled, i.e.
\begin{center}
\begin{eqnarray}
\tilde{\tilde{Z}}^{(N)}_{(ik)(jl)}&\equiv&(-)^{i+j+k+l}\text{sgn}(i-k)\text{sgn}(l-j)\nonumber\\
&&\left|\begin{array}{*{9}{c}}2p_1p_1&\ldots&2p_1p_{j-1}&2p_1p_{j+1}&\ldots&2p_1p_{l-1}
&2p_1p_{l+1}&\ldots&2p_1p_N\\\vdots&\ddots&\vdots&\vdots&\ddots&\vdots&\vdots&\ddots&\vdots\\
2p_{i-1}p_1&\ldots&2p_{i-1}p_{j-1}&2p_{i-1}p_{j+1}&\ldots&2p_{i-1}p_{l-1}&2p_{i-1}p_{l+1}&
\ldots&2p_{i-1}p_N\\2p_{i+1}p_1&\ldots&2p_{i+1}p_{j-1}&2p_{i+1}p_{j+1}&\ldots&2p_{i+1}p_{l-1}&
2p_{i+1}p_{l+1}&\ldots&2p_{i+1}p_N\\\vdots&\ddots&\vdots&\vdots&\ddots&\vdots&\vdots&\ddots&\vdots\\
2p_{k-1}p_1&\ldots&2p_{k-1}p_{j-1}&2p_{k-1}p_{j+1}&\ldots&2p_{k-1}p_{l-1}&2p_{k-1}p_{l+1}&\ldots&
2p_{k-1}p_N\\2p_{k+1}p_1&\ldots&2p_{k+1}p_{j-1}&2p_{k+1}p_{j+1}&\ldots&2p_{k+1}p_{l-1}&2p_{k+1}p_{l+1}&
\ldots&2p_{k+1}p_N\\\vdots&\ddots&\vdots&\vdots&\ddots&\vdots&\vdots&\ddots&\vdots\\
2p_Np_1&\ldots&2p_Np_{j-1}&2p_Np_{j+1}&\ldots&2p_Np_{l-1}&2p_Np_{l+1}&\ldots&2p_Np_N\end{array}\right|,\nonumber\\
&&~~~~N>2,\nonumber\\
\tilde{\tilde{Z}}^{(2)}_{(ik)(jl)}&\equiv&\delta_{il}\delta_{kj}-\delta_{ij}\delta_{kl},\nonumber\\
\hat{S}^s_{ki_2\ldots i_s}&\equiv&T^{N-1}_{(i_2)_k\ldots
(i_s)_k}(k)\bar{\delta}_{ki_2}\ldots
\bar{\delta}_{ki_s}-T^{N-1}_{i_2\ldots i_s}(0).
\end{eqnarray}
\end{center}
For $\det(Z^{(N-1)})=0$,$\det(X^{(N-1)})=0$ and all
$\tilde{X}^{(N-1)}_{0k}=0$ but $\tilde{Z}^{(N-1)}_{kl}\neq0$ and
$\tilde{X}^{N-1}_{ij}\neq0$,following equations
\begin{center}
\begin{eqnarray}
T^{N}_{{\scriptsize \underbrace{0\ldots0}_{r} \underbrace{l\ldots
l}_{n}}i_1\ldots
i_m}&=&\frac{1}{2(n+1)\tilde{Z}^{(N-1)}_{kl}}\left[-2\sum^m_{j=1}{\tilde{Z}^{(N-1)}_{ki_j}T^N_{{\scriptsize
\underbrace{0\ldots0}_{r} \underbrace{l\ldots l}_{n+1}}i_1\ldots
\hat{i}_j\ldots i_m}}\right.\nonumber\\
&&\left.+\sum^{N-1}_{j=1}{\tilde{Z}^{(N-1)}_{kj}\hat{S}^{r+n+m}_{j{\scriptsize
\underbrace{0\ldots 0}_{r-2}\underbrace{l\ldots l}_{n+1}}i_1\ldots
i_m}}\right],i_1,\ldots,i_m\neq 0,l,\nonumber\\
T^{N}_{i_1\ldots
i_s}&=&\frac{1}{\tilde{X}^{(N-1)}_{ij}}\left[\tilde{Z}^{(N-1)}_{ij}\left(2(5+s-N)T^{N}_{00i_1\ldots
i_s}-T^{N-1}_{i_1\ldots i_s}(0)\right.\right.\nonumber\\
&&\left.-\Delta L^{N-1}_{i_1\ldots i_s}-\sum^{N-1}_{n=1}{\Delta
L^{N-1}_{ni_1\ldots
i_s}}\right)\nonumber\\
&&\left.+\sum^{N-1}_{m,n=1}{\tilde{\tilde{Z}}^{(N-1)}_{(in)(jm)}f_n
\left(\hat{S}^{s+1}_{mi_1\ldots
i_s}-2\sum^s_{r=1}{\delta_{mi_r}T^N_{00i_1\ldots \hat{i}_r\ldots
i_s}}\right)}\right]
\end{eqnarray}
\end{center}
can be used. Other details in the  derivation of these equations can be found in \cite{Denner:2005nn}.

\section{Higgs Decay into Two Photons \label{sec:4}}

In this section, one-loop reduction as illustrated in the previous
section is applied to the process $H\rightarrow \gamma\gamma$.

\begin{center}
\begin{figure}
\hspace{0cm}\includegraphics[width=15cm]{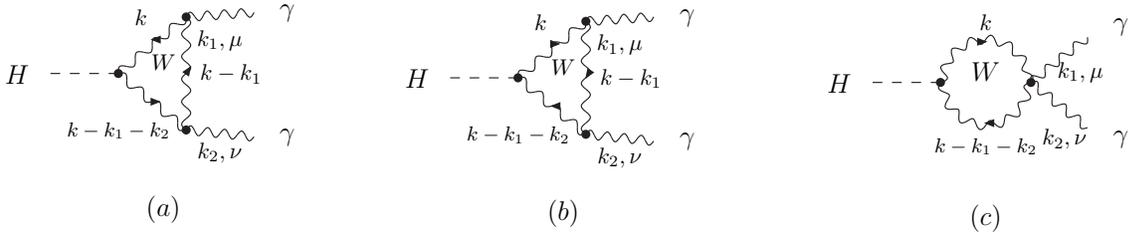}
\caption{\label{fig:w-uni} Feynman diagrams via W-boson loop in
unitary gauge for $H \rightarrow \gamma\gamma$.}
\end{figure}
\end{center}

In unitary gauge, the three diagrams via the W-boson loop that contribute
to this process with a specific loop momentum configuration are
shown in Fig.(\ref{fig:w-uni}). A direct calculation of amplitude
(dropping the polarization vectors of external photons) yields
\begin{center}
\begin{eqnarray}
\mathcal{M}^{\mu\nu}_{\text{unitary}}&=&-\frac{3e^3m_W}{8\pi^4m_H^2s_w}\left[2k_2^{\mu}k_1^{\nu}\left(i\pi^2-
(m_H^2-2m_W^2)C_0(0,0,m_H^2,m_W^2,m_W^2,m_W^2)\right)\right.\nonumber\\
&&\left.-k_1\cdot
k_2g^{\mu\nu}\left(i\pi^2-2(m_H^2-2m_W^2)C_0(0,0,m_H^2,m_W^2,m_W^2,m_W^2)\right)\right]\nonumber\\
&=&\frac{3ie^3m_W}{8\pi^2m_H^4s_w}\left[-k_2^{\mu}k_1^{\nu}\left(2m_H^2+4(m_H^2-2m_W^2)
\mathbf{f}(\frac{m_H^2}{4m_W^2})\right)\right.\nonumber\\
&&\left.+k_1\cdot
k_2g^{\mu\nu}\left(m_H^2+4(m_H^2-2m_W^2)\mathbf{f}(\frac{m_H^2}{4m_W^2})\right)\right],
\end{eqnarray}
\end{center}
with
\begin{center}
\begin{eqnarray}
\mathbf{f}(x)\equiv\left\{\begin{array}{*{3}{c}}\arcsin(\sqrt{x})^2&,&x\leq1\\
-\frac{1}{4}\left[\ln{(\frac{1+\sqrt{1-x^{-1}}}{1-\sqrt{1-x^{-1}}})}-i\pi\right]^2&,&x>1\end{array}\right.,\label{eq:1}
\end{eqnarray}
\end{center}
and the scalar integral $C_0$ is given in appendix \ref{app:b}.
However, gauge invariance is spoiled in this four-momentum cutoff
regularization. Therefore, a term should be subtracted from the
above expressions to recover gauge invariance. In this gauge, a
requirement of $\mathcal{M}^{\mu\nu}(k_1=k_2=0)=0$ should be made.
However,
\begin{center}
\begin{eqnarray}
\mathcal{M}^{\mu\nu}_{\text{unitary}}(k_1=k_2=0)&=&\frac{-3ie^3m_W}{16\pi^2s_w}g^{\mu\nu}\neq0.\label{eq:2}
\end{eqnarray}
\end{center}
Following Dyson's prescription\cite{Dyson:1949bp,Dyson:1949ha},
gauge invariance is recovered after making subtraction from
Eq.(\ref{eq:2}), and the final result is
\begin{center}
\begin{eqnarray}
\mathcal{M}^{\mu\nu}_{\text{unitary}}&=&-\frac{3ie^3}{16\pi^2m_Ws_w}\left(k_2^{\mu}k_1^{\nu}-g^{\mu\nu}k_1\cdot
k_2\right)\left(\tau^{-1}+(2\tau^{-1}-\tau^{-2})\mathbf{f}(\tau)\right)\label{eq:3}
\end{eqnarray}
\end{center}
with $\tau=\frac{m_H^2}{4m_W^2}$ following the notations of
refs.\cite{Gastmans:2011ks,Gastmans:2011wh}. Eq.(\ref{eq:3}) is the
same as those in refs.\cite{Gastmans:2011ks,Gastmans:2011wh} up to a
factor of $-2i$ from the symmetry factor of loops and different
conventions of Feynman rules. However, in this gauge, there are high
degrees of ultraviolet divergence in each diagram. The expressions
for amplitude may be different under different choices of loop
momentum. One may suspect that the discrepancy between
Eq.(\ref{eq:3}) and the result given in DREG
\begin{center}
\begin{eqnarray}
\mathcal{M}^{\mu\nu}_{\text{DREG}}&=&-\frac{ie^3}{16\pi^2m_Ws_w}\left(k_2^{\mu}k_1^{\nu}-g^{\mu\nu}k_1\cdot
k_2\right)\left(2+3\tau^{-1}+3(2\tau^{-1}-\tau^{-2})\mathbf{f}(\tau)\right)\label{eq:4}
\end{eqnarray}
\end{center}
is originated from the bad loop momentum choices in Eq.(\ref{eq:3}).
However, in our  calculation we find that the terms
\begin{center}
\begin{eqnarray}
\Delta
\mathcal{M}^{\mu\nu}(p)&=&-\frac{ie^3}{96\pi^2m_W^3s_w}\left[\left(k_2^{\mu}p^{\nu}-p^{\mu}k_1^{\nu}\right)
\left(-3\Lambda^2-2m_H^2-6m_W^2+(k_1+k_2)\cdot
p-p^2\right)\right.\nonumber\\&& \left.+2g^{\mu\nu}(k_1-k_2)\cdot
p\left(-3\Lambda^2-2m_H^2+3m_W^2+(k_1+k_2)\cdot
p-p^2\right)\right]\label{eq:5}
\end{eqnarray}
\end{center}
should be added to Eq.(\ref{eq:3}) if loop momentum $k$ is shifted
to $k+p$. From the symmetric consideration of $k_1,k_2$, a proper
choice of $p$ is $\frac{k_1+k_2}{2}$  which is the same as that
presented in refs.\cite{Gastmans:2011ks,Gastmans:2011wh}. Since
$\Delta \mathcal{M}^{\mu\nu}(\frac{k_1+k_2}{2})=0$ in
Eq.(\ref{eq:5}), the result in Eq.(\ref{eq:3}) remains unchanged.
From Eq.(\ref{eq:5}), it seems hopeless that the difference between
Eq.(\ref{eq:3}) and Eq.(\ref{eq:4}) can be eliminated through
shifting the integral momentum $k$.

\begin{center}
\begin{figure}
\hspace{0cm}\includegraphics[width=15cm]{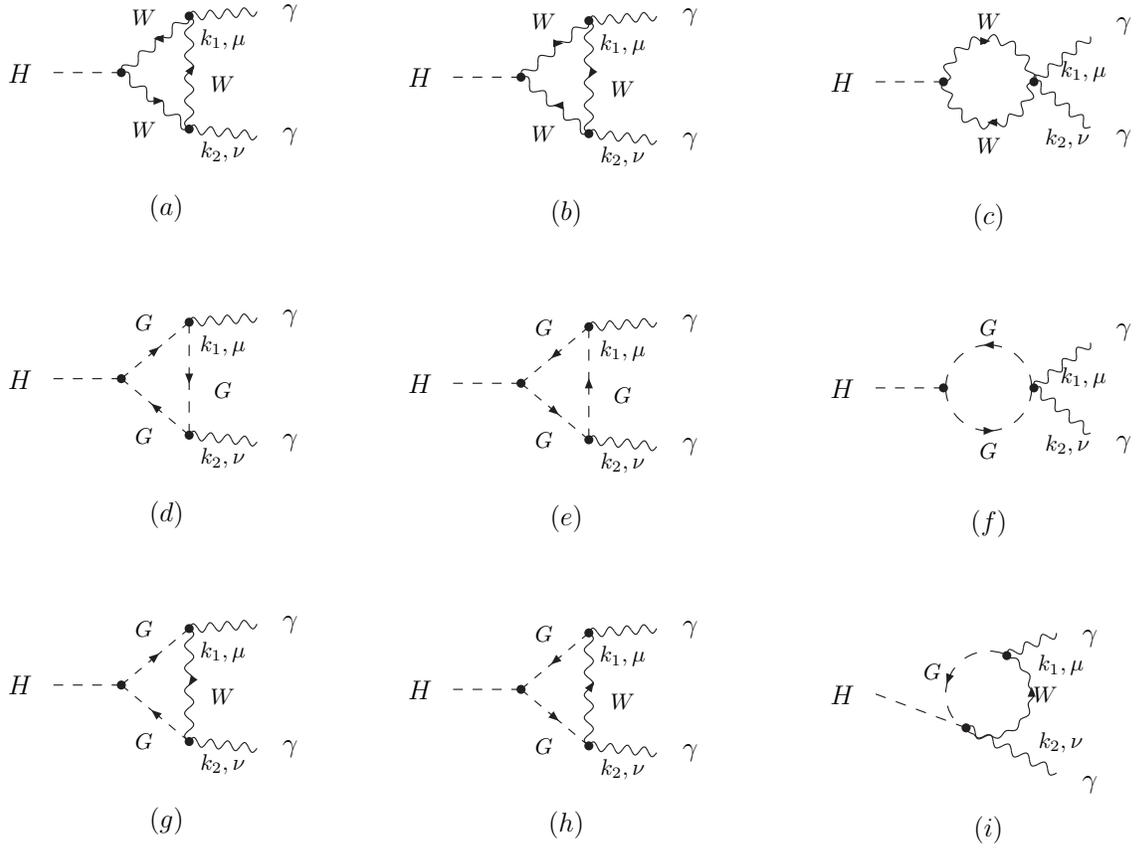}
\caption{\label{fig:w-fey} Some representative Feynman diagrams via
W-boson loop in 't Hooft-Feynman gauge for $H \rightarrow
\gamma\gamma$.}
\end{figure}
\end{center}

In 't Hooft-Feynman gauge ($\xi=1$), the amplitude with one-loop
diagrams shown in Fig.(\ref{fig:w-fey}) is
\begin{center}
\begin{eqnarray}
\mathcal{M}^{\mu\nu}_{\xi=1}&=&\frac{e^3}{16\pi^4m_H^2m_Ws_w}\left\{2k_2^{\mu}k_1^{\nu}
\left[-i\pi^2(m_H^2+6m_W^2)\right.\right.\nonumber\\
&&\left.+6m_W^2(m_H^2-2m_W^2)C_0(0,0,m_H^2,m_W^2,m_W^2,m_W^2)\right]
+k_1\cdot
k_2g^{\mu\nu}\left[i\pi^2(m_H^2+6m_W^2)\right.\nonumber\\&&\left.\left.
-12m_W^2\left(m_H^2-2m_W^2\right)C_0(0,0,m_H^2,m_W^2,m_W^2,m_W^2)\right]\right\}\nonumber\\
&=&\frac{ie^3}{16\pi^2m_H^2m_Ws_w}\left[-2k_1^{\nu}k_2^{\mu}\left(m_H^4+6m_H^2m_W^2+12m_W^2
\left(m_H^2-2m_W^2\right)\mathbf{f}(\tau)\right)\right.\nonumber\\
&&\left.+k_1\cdot
k_2g^{\mu\nu}\left(m_H^4+6m_H^2m_W^2+24m_W^2(m_H^2-2m_W^2)\mathbf{f}(\tau)\right)\right].\label{eq:6}
\end{eqnarray}
\end{center}
Following a similar procedure to obtain a gauge invariant result,
the amplitude at $k_1=k_2=0$ is calculated as
\begin{center}
\begin{eqnarray}
\mathcal{M}^{\mu\nu}_{\xi=1}(k_1=k_2=0)&=&-\frac{3ie^3m_W}{16\pi^2s_w}g^{\mu\nu},
\end{eqnarray}
\end{center}
which is the same as Eq.(\ref{eq:3}). However, the gauge invariant
amplitude is non-vanishing at $k_1=k_2=0$ because of the
contributions of diagrams $(g)$ and $(h)$ in Fig.(\ref{fig:w-fey})
in this gauge. These contributions are
\begin{center}
\begin{eqnarray}
\mathcal{M}^{\mu\nu}_{\xi=1,(g,h)}(k_1=k_2=0)&=&\frac{ie^3m_H^2}{32\pi^2m_Ws_w}g^{\mu\nu}.
\end{eqnarray}
\end{center}
Therefore, the subtracted terms should be
$\mathcal{M}^{\mu\nu}_{\xi=1}(k_1=k_2=0)-\mathcal{M}^{\mu\nu}_{\xi=1,(g,h)}(k_1=k_2=0)$
instead of $\mathcal{M}^{\mu\nu}_{\xi=1}(k_1=k_2=0)$. The final
result is
\begin{center}
\begin{eqnarray}
\mathcal{M}^{\mu\nu}_{\xi=1}&=&-\frac{ie^3}{16\pi^2m_Ws_w}\left(k_2^{\mu}k_1^{\nu}-g^{\mu\nu}k_1\cdot
k_2\right)\left(2+3\tau^{-1}+3(2\tau^{-1}-\tau^{-2})\mathbf{f}(\tau)\right)\nonumber\\
&=&\mathcal{M}^{\mu\nu}_{\text{DREG}}.\label{eq:7}
\end{eqnarray}
\end{center}
The term generated by the contributions of the Goldstone triangle
diagrams $(d,e)$ in Fig.(\ref{fig:w-fey}) spoils the decoupling
theorem, as pointed out by Shifman {\it et al.} recently
\cite{Shifman:2011ri}. Given that there are only logarithmic
divergences under this covariant gauge, the result in
Eq.(\ref{eq:7}) is unique with a different loop momentum chosen. To
best of our knowledge, it is the first derivation in the 't
Hooft-Feynman gauge in cutoff regularization.

\begin{center}
\begin{figure}
\hspace{0cm}\includegraphics[width=10cm]{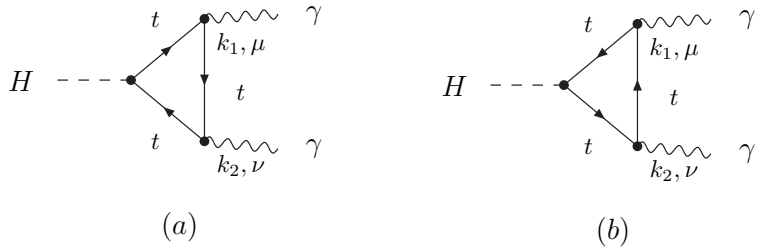}
\caption{\label{fig:top} Feynman diagrams via top-quark loop for $H
\rightarrow \gamma\gamma$.}
\end{figure}
\end{center}

It seems there are some problems with unitary gauge in this cutoff
regularization. Given that the top quark loop ( Fig.(\ref{fig:top}))
does not suffer from any ambiguities in the gauge or loop momentum
choices, the diagrammatic expressions are expected to be the same in
DREG and in this cutoff regularization. These conditions have been
verified  following the same procedures. The result is as follows
\begin{center}
\begin{eqnarray}
\mathcal{M}^{\mu\nu}_{\text{top}}&=&\frac{ie^3N_c}{18\pi^2m_Ws_w}\left(k_2^{\mu}k_1^{\nu}-k_1\cdot
k_2g^{\mu\nu}\right)\left(\chi^{-1}+(\chi^{-1}-\chi^{-2})\mathbf{f}(\chi)\right),
\end{eqnarray}
\end{center}
where $\chi=\frac{m_H^2}{4m_t^2}$.

The authors of refs.\cite{Huang:2011yf,Marciano:2011} have also
calculated this Higgs decay process in Pauli-Villars regularization
and dimensional regularization respectively, and obtained the same
result as the old
ones\cite{Ellis:1975ap,Ioffe:1976sd,Shifman:1979eb,Rizzo:1979mf}.
Their statement about this issue is that the integral(in Euclidean)
\begin{center}
\begin{eqnarray}
I_{\mu\nu}&\equiv&\int{\frac{k^2g_{\mu\nu}-4k_{\mu}
k_{\nu}}{(k^2+m^2)^3}}
\end{eqnarray}
\end{center}
is vanishing in cutoff regularization, while it is nonzero in DREG,
which is also pointed out by R.Gastmans
\cite{Gastmans:2011ks,Gastmans:2011wh}. They also argued that
$I_{\mu\nu}$ contained the difference of two logarithmic
divergencies and should be regulated. Therefore, the integral that
violats electromagnetic gauge invariance may suffer from some
ambiguities. Actually, this issue was first discussed by
R.Jackiw\cite{Jackiw:1999qq} in a more general case. However,we
think that the vanishing of $I_{\mu\nu}$ in 4 dimensions is just a
result of the fact that the integral intervals are symmetric about
the origin even when there is a cutoff $\mathbf{\Lambda}$, and a
replacement of $k_{\mu}k_{\nu}\rightarrow\frac{g_{\mu\nu}k^2}{4}$ in
the integrand is also proper.

Moreover, very recently R.Jackiw also pointed out that by combining
the two terms in the integrand of $I_{\mu\nu}$ one can avoid
infinities but the difference of the integrals in these two
regularization schemes is still the same, thus both evaluations are
mathematically defensible\cite{Jackiw:2011}. So, what are the
physical reasons for these ambiguities? In the following, we will
try to clarify this issue.

Considering that the diagrammatic expressions in unitary gauge are
not well-defined in the four-momentum cutoff regularization, how to
recover the correct result under this condition may be still an open
question. We investigate the Lagrangian for the Standard Model in
unitary gauge, similar to the treatments in
ref.\cite{Sonoda:2001um}. The covariant terms for scalars are
\begin{center}
\begin{eqnarray}
\mathfrak{L}_{\text{scalar}}&\equiv&({\rm
D}_{\mu}\Phi)^{\dagger}({\rm
D}^{\mu}\Phi)-{\rm V}(\Phi),\nonumber\\
\text{with}~~{\rm
V}(\Phi)&\equiv&-\mu^2\Phi^{\dagger}\Phi+\lambda(\Phi^{\dagger}\Phi)^2,
{\rm D}_{\mu}\Phi\equiv\left(\partial_{\mu}-\frac{i}{2}{\rm
g}\tau^i{\rm W}^i_{\mu}-\frac{i}{2}{\rm g}^{\prime}{\rm
B}_{\mu}\right)\Phi.\label{eq:8}
\end{eqnarray}
\end{center}
The scalar doublet produces the vacuum expectation value through the Higgs
mechanism as
\begin{center}
\begin{eqnarray}
\langle \Phi \rangle_0&=&\left(\begin{array}{c}0\\\frac{{\rm
v}}{\sqrt{2}}\end{array}\right),{\rm
v}=\left(\frac{\mu^2}{\lambda}\right)^{1/2}.
\end{eqnarray}
\end{center}
Therefore, the scalar fields can be redefined as
\begin{center}
\begin{eqnarray}
\Phi&=&\left(\begin{array}{c}\phi^+\\ \frac{{\rm v}}{\sqrt{2}}+{\rm
h}+i\phi_3\end{array}\right).
\end{eqnarray}
\end{center}
There are terms like
\begin{center}
\begin{eqnarray}
&&m_W^2{\rm W}^+_{\mu}{\rm W}^{-,\mu}+i m_W\left({\rm
W}^-_{\mu}\partial^{\mu}{\phi^+}-{\rm
W}^+_{\mu}\partial^{\mu}{\phi^-}\right)\nonumber\\
&=&m_W^2\left({\rm
W}^+_{\mu}+\frac{i}{m_W}\partial_{\mu}{\phi^+}\right)\left({\rm
W}^{-,\mu}-\frac{i}{m_W}\partial^{\mu}{\phi^-}\right)-\partial_{\mu}{\phi^+}\partial^{\mu}{\phi^-}\label{eq:9}
\end{eqnarray}
\end{center}
after expanding the Lagrangian given in Eq.(\ref{eq:8}), where
$W^{\pm}_{\mu}\equiv\frac{W^1_{\mu}\mp i W^2_{\mu}}{\sqrt{2}}$. By
following the prescription in ref.\cite{Sonoda:2001um}, the W-boson fields in unitary gauge can be redefined as
\begin{center}
\begin{eqnarray}
\tilde{{\rm W}}^+_{\mu}&\equiv&{\rm
W}^+_{\mu}+\frac{i}{m_W}\partial_{\mu}{\phi^+},\nonumber\\
\tilde{{\rm W}}^-_{\mu}&\equiv&{\rm
W}^-_{\mu}-\frac{i}{m_W}\partial_{\mu}{\phi^-}.\label{eq:10}
\end{eqnarray}
\end{center}
In this gauge there are no kinetic term
$\partial_{\mu}{\phi^+}\partial^{\mu}{\phi^-}$ and mass term for the
Goldstone $\phi^+,\phi^-$ because of the cancelation between the
last term in Eq.(\ref{eq:9}) and the original kinetic term of the
W-boson's Goldstone in $\mathfrak{L}_{\text{scalar}}$. However,
terms such as $h\phi^+\phi^-$ still exist in the original
Lagrangian. In DREG, $\phi^+=\phi^-=0$ can be set safely, similar to
a previous work by Grosse-Knetter \cite{GrosseKnetter:1992nn},
because all the momentum modes can be included in this
regularization\footnote{Note that the limits of loop integrals are
taken to be infinity}. Hence, the conventional Lagrangian in unitary
gauge only with physical fields is obtained. However, the results
are in contrast to those of the four-momentum cutoff regularization,
because an artificial scale $\mathbf{\Lambda}$ is introduced in the
Lagrangian. The absence of a kinetic term for $\phi^+,\phi^-$ does
not mean that these Goldstone fields are vanishing intuitively, but
because the theory does not provide any information above
$\mathbf{\Lambda}$ in this regularization. If the mass of
$\phi^+,\phi^-$ is assumed to be $\mathcal{O}(\mathbf{\Lambda})$,
there are still finite contributions from the Goldstone triangle
diagrams when $\mathbf{\Lambda}\rightarrow \infty$. From this
viewpoint, the cutoff regularization in unitary gauge is
problematic. The violation of the property in gauge invariance can
also be attributed to the absence of large momentum modes.
Therefore, the old results in the literature for $H\rightarrow
\gamma\gamma$ are still valid.

In fact, dimensional and Pauli-Villars regularization schemes are
free of missing large momentum modes, and can maintain gauge
invariance. Therefore, these results are correct also in unitary
gauge. From the evaluations of R.Jackiw, the integral $I_{\mu\nu}$
can be dealt without any infinities, and the only difference is from
surface terms (i.e.,large momentum region), which also verifies our
conclusion.

\section{Summary \label{sec:5}}
A method for systematical evaluations of one-loop tensor integrals
in cutoff regularization is proposed by deriving a new recursive
relation Eq.(\ref{eq:In3}) and implementing it in the
Passarino-Veltman reduction method. The result has been expressed in
a form that can be directly translated into computer codes. Similar
to the methods presented in ref.\cite{Denner:2005nn}, our results
are also numerical stable for up to four-point integrals. Surely,
our method can be extended to deal with high-point integrals
straightforwardly.

With this approach, we have calculated the amplitudes for Higgs
decay into two photons via the W-boson loop and the top-quark loop.
The correctness of the method has been confirmed by evaluating these
processes and checking other programs, and it is certainly useful in
both theoretical and phenomenological aspects. Moreover, we also
reanalyze the Higgs decay process and make our efforts to find the
physical reasons for some puzzles appeared in the calculations of
this process.

\begin{acknowledgments}
We thank R.Jackiw for providing us with his evaluations for some
loop integrals. This work was supported by the National Natural
Science Foundation of China (No 10805002, 10847001, 11021092,
11075002, 11075011), the Foundation for the Author of National
Excellent Doctoral Dissertation of China (Grant No. 201020), and the
Ministry of Science and Technology of China (2009CB825200).

\end{acknowledgments}

\begin{appendix}

\section{Derivation of Expressions for $J^N_{\mu_1\mu_2...\mu_s}$ \label{app:a}}

In this appendix, the general formulae for
$J^N_{\mu_1\mu_2...\mu_s}\equiv\int{{\rm
d}^4k\frac{k_{\mu_1}k_{\mu_2}...k_{\mu_s}}{(-k^2+a^2)^N}}$ used in
section \ref{sec:2} are derived first. Obviously,
$J^N_{\mu_1\mu_2...\mu_s}$ is vanishing until $\mathbf{s}$ is even.
Therefore, $\mathbf{s}$ should be an even and non-negative integer
and $N$ should be positive in the following context.

A notation (similar to but a little different from that in
Ref.\cite{Denner:2005nn}) is introduced first in order to write down
the tensor decomposition in a concise way. We use curly braces to
denote symmetrization with respect to Lorentz indices, where all
non-equivalent permutations of the Lorentz indices on metric tensor
$\mathbf{g}$ and momenta $p$ are implicitly understood. A generic
notation $\{g^{2n_0}p_1^{n_1}\ldots
p_k^{n_k}\}_{\mu_1\ldots\mu_{t}}$ with $t=\sum^{k}_{l=1}{n_l}+2n_0$
means a sum that the $\mathbf{2n_0}$ of Lorentz indices
$\mu_1,\ldots,\mu_t$ are distributed to $\mathbf{n_0}$ metric
tensors $\mathbf{g}$ while $\mathbf{n_l}$ of them are distributed to
$\mathbf{n_l}$ momenta $\mathbf{p_l}$ with equal weights. For
instance,
\begin{center}
\begin{eqnarray}
\{g^4\}_{\mu\nu\rho\sigma}&\equiv&
g_{\mu\nu}g_{\rho\sigma}+g_{\mu\rho}g_{\nu\sigma}+g_{\mu\sigma}g_{\nu\rho},\nonumber\\
\{g^2p^1\}_{\mu\nu\rho}&\equiv&
g_{\mu\nu}p_{\rho}+g_{\nu\rho}p_{\mu}+g_{\rho\mu}p_{\nu},\nonumber\\
\{p^2_1p^1_2\}_{\mu\nu\rho}&\equiv&p_{1\mu}p_{1\nu}p_{2\rho}+p_{1\mu}p_{1\rho}p_{2\nu}+p_{1\nu}p_{1\rho}p_{2\mu}.
\end{eqnarray}
\end{center}

The Lorentz covariance ensures us to make the following replacement
\begin{center}
\begin{eqnarray}
k_{\mu_1}k_{\mu_2}...k_{\mu_s} \longrightarrow
\{g^s\}_{\mu_1\mu_2...\mu_s}\frac{(k^2)^\frac{s}{2}}{\Gamma(\frac{s}{2}+2)2^{\frac{s}{2}}}
\end{eqnarray}
\end{center}
in the integral $J^N_{\mu_1...\mu_s}$, which can be proven by the
induction of the integer $\mathbf{s}$. After this replacement and
subsequent Wick rotation, spherical coordinate system transformation
and some trivial variable substitutions, one arrived
\begin{center}
\begin{eqnarray}
J^N_{\mu_1...\mu_s}&=&\{g^s\}_{\mu_1...\mu_s}\frac{i\pi^2(-2)^{-\frac{s}{2}}}{\Gamma(\frac{s}{2}+2)}
\int_{0}^{\Lambda^2}{{\rm d}K\frac{K^\frac{s+2}{2}}{(K+a^2)^N}},
\label{eq:Jn}
\end{eqnarray}
\end{center}
where $\mathbf{\Lambda}$ was denoted as the ultraviolet cutoff
scale.

Eq.(\ref{eq:Jn}) can be solved directly when $a^2=0$, i.e.
\begin{center}
\begin{eqnarray}
J^N_{\mu_1...\mu_s}&=&\{g^s\}_{\mu_1...\mu_s}\frac{i\pi^2(-2)^{-\frac{s}{2}}}
{\Gamma(\frac{s}{2}+2)}\frac{2\Lambda^\Delta}{\Delta}
\end{eqnarray}
\end{center}
when superficial degree of ultraviolet divergence $\Delta\equiv
s-2N+4>0$. In the case of $\Delta\leq0$, Eq.(\ref{eq:Jn}) encounters
infrared divergence, which is not considered in this article. When
$a^2\neq0$, result becomes a little more complicated than the
previous one. However, this problem can be resolved after
implementing the tricks of using integration by parts and following
integral formulae
\begin{center}
\begin{eqnarray}
\int{{\rm d}x~x^n\ln(x+a)}&=&\frac{1}{n+1}\left(x^{n+1}\ln(x+a)
-\sum_{k=0}^{n}{\frac{(-a)^kx^{n+1-k}}{n+1-k}}\right.\nonumber\\&&
\left.-(-a)^{n+1}\ln(a+x)\right),~~~~~~~~~~~~~n\in \mathbf{N}
\end{eqnarray}
\end{center}
into Eq.(\ref{eq:Jn}). The explicit expressions for
$J^N_{\mu_1\ldots\mu_s}$ can also be obtained, i.e. $a^2\neq0$ and
superficial degree of divergence $\Delta\equiv s-2N+4\geq0$ yields
\begin{center}
\begin{eqnarray}
J^N_{\mu_1...\mu_s}&=&\{g^s\}_{\mu_1...\mu_s}~(-2)^{-\frac{s}{2}}\frac{i\pi^2}{\Gamma(N)}\nonumber\\
&&\left[-\sum_{k=1}^{N-1}{\frac{\Gamma(N-k)}{\Gamma(\frac{s}{2}+3-k)}
\left(\sum_{l=0}^{\frac{\Delta}{2}}{C^{l}_{N-k+l-1}(-a^2)^l\Lambda^{\Delta-2l}}\right)}\right.\nonumber\\
&&\left.+\frac{\Gamma(1)}{\Gamma(\frac{\Delta}{2}+1)}\left(
\sum_{k=0}^{\frac{\Delta}{2}-1}{(-a^2)^{k}\frac{2\Lambda^{\Delta-2k}}{\Delta-2k}}
~+(-a^2)^{\frac{\Delta}{2}}\ln(\frac{\Lambda^2}{a^2})\right)\right],
\end{eqnarray}
\end{center}
while $a^2\neq0$ but $\Delta<0$ returns
\begin{center}
\begin{eqnarray}
J^N_{\mu_1...\mu_s}&=&\{g^s\}_{\mu_1...\mu_s}~(-2)^{-\frac{s}{2}}\frac{i\pi^2\Gamma(-\frac{\Delta}{2})}{\Gamma(N)}a^\Delta.
\end{eqnarray}
\end{center}

\section{Some Scalar Integrals \label{app:b}}

After the reduction of one-loop integrals using the modified
Passarino-Veltman reduction formulas given in the section
\ref{sec:2}, every tensor integral can be expressed as a linear
combination of up to four-point scalar integrals. In this appendix,
the analytical expressions for some scalar integrals are listed
below. Some of them may be used in the one-loop calculations of the
process Higgs decay to two photons.

First of all, the conventions for the scalar integrals used
in this article are fixed as follows
\begin{center}
\begin{eqnarray}
T^N_0&\equiv&\int{{\rm d}^4k\frac{1}{D_0D_1\ldots D_{N-1}}}.
\end{eqnarray}
\end{center}

For one-point functions,
\begin{center}
\begin{eqnarray}
A_0(0)&=&-i\pi^2~\Lambda^2,\nonumber\\
A_0(m_0^2)&=&i\pi^2m_0^2\left(\ln(\frac{\Lambda^2}{m_0^2})-\frac{\Lambda^2}{m_0^2}\right),\nonumber\\
A_{\scriptsize \underbrace{0\ldots
0}_{2n}}(m_0^2)&=&\frac{(-)^{n+1}i\pi^2}{\Gamma(n+2)2^n}\left(\sum^{n+1}_{i=1}{\frac{(-)^{n+1-i}}{i}
\Lambda^{2i}m_0^{2n+2-2i}}+(-)^{n+1}m_0^{2n+2}\ln(\frac{\Lambda^2}{m_0^2})\right).
\end{eqnarray}
\end{center}
Two-point functions can be easily verified as
\begin{center}
\begin{eqnarray}
B_0(p_1^2,m_0^2,m_1^2)&=&i\pi^2\left(\ln(\frac{\Lambda^2}{p_1^2})+1+\sum^2_{i=1}{[\gamma_i
\ln(\frac{\gamma_i-1}{\gamma_i})-\ln(\gamma_i-1)]}\right),\nonumber\\
\text{with}~~\gamma_{1,2}&=&\frac{p_1^2-m_1^2+m_0^2\pm\sqrt{\left(p_1^2-m_1^2+m_0^2\right)^2-4p_1^2m_0^2}}{2p_1^2},\nonumber\\
B_0(0,0,m^2)&=&i\pi^2\ln(\frac{\Lambda^2}{m^2}),\nonumber\\
B_0(p^2,0,0)&=&i\pi^2\left(\ln(\frac{\Lambda^2}{p^2})+1\right),\nonumber\\
B_0(p^2,0,m^2)&=&i\pi^2\left(\ln(\frac{\Lambda^2}{m^2})+1+\frac{m^2-p^2}{p^2}\ln(\frac{m^2-p^2}{m^2})\right),\nonumber
\end{eqnarray}
\end{center}
\begin{center}
\begin{eqnarray}
B_0(0,m_0^2,m_1^2)&=&i\pi^2\frac{m_0^2\ln(\frac{\Lambda^2}{m_0^2})-m_1^2\ln(\frac{\Lambda^2}{m_1^2})}{m_0^2-m_1^2},\nonumber\\
B_0(0,m^2,m^2)&=&i\pi^2\left(\ln(\frac{\Lambda^2}{m^2})-1\right).
\end{eqnarray}
\end{center}
Moreover, two special finite three-point functions are
\begin{center}
\begin{eqnarray}
C_0(0,0,p^2,m^2,m^2,m^2)&=&\frac{i\pi^2}{2p^2}\left(\ln^2(\frac{1-\sqrt{1-\frac{4m^2}{p^2}}}
{1+\sqrt{1-\frac{4m^2}{p^2}}})-\pi^2\right),\nonumber\\
C_0(0,0,0,m^2,m^2,m^2)&=&-\frac{i\pi^2}{2m^2}.
\end{eqnarray}
\end{center}
\end{appendix}




\providecommand{\href}[2]{#2}\begingroup\raggedright\endgroup

\end{document}